\documentclass[12pt]{article}

\usepackage{amsfonts,amssymb,amsmath}
\usepackage{graphicx}
\DeclareGraphicsExtensions{.pdf,.png,.jpg}
\begin{document}
\title{\bf Controlling speedup in open quantum systems through the manipulating of system-reservoir bound states}
\author{ N. Behzadi $^{a}$
\thanks{E-mail:n.behzadi@tabrizu.ac.ir}  ,
B. Ahansaz  $^{b}$,
A. Ektesabi $^{b}$ and
E. Faizi $^{b}$
\\ $^a${\small Research Institute for Fundamental Sciences, University of Tabriz, Tabriz, Iran,}
\\ $^b${\small Physics Department, Azarbaijan Shahid Madani University, Tabriz, Iran,}}\maketitle

\begin{abstract}
In this paper, we give a mechanism for controlling speedup of a single-qubit open
quantum system by exclusively manipulating the system-reservoir bound states using
additional non-interacting qubits. It is demonstrated that providing stronger bound
states in the system-reservoir spectrum makes the single-qubit evolution with higher
speed. We examine the performance of the mechanism for different spectral densities
such as Lorentzian and Ohmic and find out the decisive role of bound states manipulation in
speeding up of quantum evolution.
\\
\\
{\bf PACS Nos:}
\end{abstract}

\section{Introduction}
Quantum mechanics imposes a fundamental limit to the speed of quantum evolution, conventionally known as quantum speed limit (QSL) time which is the minimum evolution time for a quantum state to become a different state. The QSL plays an important role in many areas of quantum physics, such as quantum computation $\cite{Bekenstein,Lloyd}$, quantum metrology [3-5], optimal control $\cite{Caneva}$ and quantum communication $\cite{Duan,Yung}$. For closed systems, two types of QSL time bound have been derived: the Mandelstam-Tamm (MT) bound $\cite{Mandelstam}$ $\tau_{QSL}= \pi\hbar/(2\Delta E)$ and Margolus-Levitin $\cite{Margolus}$ bounds $\tau_{QSL}= \pi\hbar/(2E)$, where $\Delta E$ is the variance of energy and $E$ its average with respect to the ground state. Both the Mandelstamm-Tamm and Margolus-Levitin bounds are attainable in closed quantum systems for initial pure states. Since any system is inevitably subjected to an environment, QSL time bound for open quantum systems is highly desirable. Taddei $et$ $al$. $\cite{Taddei}$ extended the Mandelstamm-Tamm type bound to both unitary and nonunitary processes described by positive non-unitary maps by using of quantum Fisher information for time estimation. Later Deffner $et$ $al$. $\cite{Deffner}$ extended both Mandelstamm-Tamm and Margolus-Levitin bounds for nonunitary evolution of open quantum system by using the geometric approach provided by the Bures angle $\cite{Bures}$. However this bound is attainable for initially pure states and it is not feasible for mixed states. The QSL time bound for both pure and mixed initial states has been derived by employing the relative purity $\cite{del Campo}$ and different types of alternative fidelity $\cite{Zhe Sun,pra}$. In all of them it was showen that the non-Markovian effects can speedup quantum evolution and subsequently leads to smaller QSL time bound. The authors in Ref. $\cite{Bin}$ showed that the appearance of non-Markovianity, as a resource of quantum speedup, in the dynamics of a two-level system is related to the formation of bound state in the whole system spectrum.

In this paper, we investigate a mechanism for controllable speeding up of evolution of a single-qubit open quantum system  exclusively on the basis of manipulating system-reservoir bound states using additional non-interacting qubits. We consider a system consisting of $N$ non-interacting qubits in a dissipative common reservoir and reveal that, increasing $N$ leads to the stronger bound states for the system-reservoir. Our results suggests that the stronger bound states provided by adding $N-1$ non-interacting qubits, can accelerate the quantum evolution of the single-qubit system. We also examine the performance
of the mechanism for reservoirs with different spectral densities such as Lorentzian and Ohmic, and
observe that the speedup process for the single-qubit evolution depends only on the
manipulating of system-reservoir bound states.

The work is organized as follows. In Sec. II, the dynamical model of $N$ non-interacting qubits in a common zero-temperature reservoir is introduced. In Sec. III, we derive the condition for the formation of bound states
for the qubits and the related reservoir by exploring the eigen-spectrum of the model. Sec. IV is devoted to derive a QSL time bound for a given single-qubit in this system. In Sec. V, we present the role of additional qubits in speeding up of a single-qubit evolution in the framework of QSL time for a single-qubit system by considering Lorentzian and Ohmic spectral density for the reservoir. Finally, the paper is ended by a brief conclusion.

\section{Dynamics of single-qubit open system in the presence of $N-1$ additional qubits}

We consider a single-qubit system along with $N-1$ similar non-interacting qubits (two-level atoms) involved totally in a common zero-temperature thermal reservoir.
The Hamiltonian of the whole system is given by
\begin{eqnarray}
\hat{H}=\hat{H}_{0}+\hat{H}_{I},
\end{eqnarray}
where
\begin{eqnarray}
\hat{H}_{0}=\omega_{0} \sum_{l=1}^{N} \hat{\sigma}^{+}_{l} \hat{\sigma}^{-}_{l}+\sum_{k} \omega_{k} \hat{b_{k}}^{\dagger} \hat{b_{k}},
\end{eqnarray}
and
\begin{eqnarray}
\hat{H}_{I}=\sum_{l=1}^{N} \sum_{k} g_{k} \hat{b_{k}} \hat{\sigma}_{l}^{+}+g_{k}^{*} \hat{b_{k}^{\dagger}} \hat{\sigma}_{l}^{-},
\end{eqnarray}
in which $\hat{\sigma}_{l}^{+}(\hat{\sigma}_{l}^{-})$ is the raising (lowering) operator of the $l^{th}$ qubit with transition frequency $\omega_{0}$
and $\hat{b_{k}}$ ($\hat{b_{k}^{\dagger}}$) is the annihilation (creation) operator of the $k^{th}$ field mode with frequency $\omega_{k}$. Also, the strength of coupling between the $l^{th}$ qubit and the $k^{th}$ field mode is represented by $g_{k}$. In the interaction picture the Schrodinger equation is written as
\begin{eqnarray}
  i \frac{d}{dt}|\psi(t)\rangle=\hat{H}_{I}(t) |\psi(t)\rangle,
\end{eqnarray}
where the interaction term (3), in this picture, is given by
\begin{eqnarray}
\hat{H}_{I}(t)=e^{i \hat{H}_{0}t} \hat{H}_{I} e^{-i \hat{H}_{0}t}=\sum_{l=1}^{N} \sum_{k} g_{k} {\hat{\sigma}_{l}^{+}} \hat{b_{k}} e^{i(\omega_{0}-\omega_{k})t}+g_{k}^{*} {\hat{\sigma}_{l}^{-}} \hat{b_{k}}^{\dagger} e^{-i(\omega_{0}-\omega_{k})t}.
\end{eqnarray}
We can immediately see that the total Hamiltonian commutes with the number of excitations, i.e.
\begin{eqnarray}
 \Big[\sum_{l=1}^{N} {\hat{\sigma}_{l}^{+} \hat{\sigma}_{l}^{-}}+\sum_{k} \hat{b_{k}}^{\dagger} \hat{b_{k}}, H\Big]=0.
\end{eqnarray}
Therefore by considering the single excitation subspace for the whole system, the initial state is assumed to be as follows
\begin{eqnarray}
  |\psi(0)\rangle=C_{0}(0)|0\rangle_{S} |0\rangle_{E}+\sum_{l=1}^{N}C_{l}(0)|l\rangle_{S} |0\rangle_{E}.
\end{eqnarray}
After time $t>0$, the state (7) evolves to the following one
\begin{eqnarray}
\begin{array}{c}
  |\psi(t)\rangle=C_{0}(t)|0\rangle_{S} |0\rangle_{E}+\sum_{l=1}^{N}C_{l}(t)|l\rangle_{S} |0\rangle_{E}+\sum_{k} C_{k}(t) |0\rangle_{S} |1_{k}\rangle_{E},
\end{array}
\end{eqnarray}
where $|l\rangle_{S}=|g\rangle^{\bigotimes N}_{l^{th}\equiv e}$ means that all of the qubits are in the ground state $|g\rangle$ except the $l^{th}$ qubit
which is in the excited state $|e\rangle$ and $|0\rangle_{S}=|g\rangle^{\bigotimes N}=|g,g,...,g\rangle$.
Also, we denote $|0\rangle_{E}$ being the vacuum state of the reservoir and $|1_{k}\rangle_{E}$ is the state for which there is only one excitation in the $k^{th}$ field mode. From Eq. (4), it is clear that $\dot{C}_{0}(t)=0$, then we have $C_{0}(t)=C_{0}(0)=C_{0}$ and the other probability amplitudes are given by the following integro-differential equations
\begin{eqnarray}
  \frac{dC_{l}(t)}{dt}=-\int_{0}^{t} f(t-t')\sum_{m=1}^{N}C_{m}(t') dt',
\end{eqnarray}
where $l=1,2,...,N$, and the correlation function $f(t-t')$ is related to the spectral density $J(\omega)$ of the reservoir by
\begin{eqnarray}
f(t-t')=\int d\omega J(\omega) e^{i(\omega_{0}-\omega)(t-t')}.
\end{eqnarray}
The exact form of $C_{l}(t)$s thus depend on the particular choice of the spectral density of the reservoir.
In the next sections, the first qubit ($l=1$) is considered as our main concern of single-qubit system and the $N-1$ remainder ones are considered as the additional qubits.

\section{Formation of bound states for the total system}

The energy spectrum of the whole system discussed in the previous section can be obtained by solving the eigenvalue equation
\begin{eqnarray}
 \mathcal{\hat{H}}(t) |\psi(t)\rangle=E |\psi(t)\rangle,
\end{eqnarray}
where $\mathcal{\hat{H}}(t)=e^{i \hat{H}_{0}t} \hat{H} e^{-i \hat{H}_{0}t}=\hat{H}_{0}+\hat{H}_{I}(t)$ is the Hamiltonian of the total system in the interaction picture and $|\psi(t)\rangle$ is given as Eq. (8).
Eq. (11) imposes that $C_{0}=0$ and also yields the following set of $N+1$ equations
\begin{eqnarray}
\begin{array}{c}
  \omega_{k} C_{k}(t)+\sum_{l=1}^{N} g_{k}^{*} e^{-i(\omega_{0}-\omega_{k})t} C_{l}(t)=E C_{k}(t),\\\\
  \omega_{0} C_{1}(t)+\sum_{k} g_{k} e^{i(\omega_{0}-\omega_{k})t} C_{k}(t)=E C_{1}(t),\\\\
  \omega_{0} C_{2}(t)+\sum_{k} g_{k} e^{i(\omega_{0}-\omega_{k})t} C_{k}(t)=E C_{2}(t),\\
  .\\
  .\\
  .\\
  \omega_{0} C_{N}(t)+\sum_{k} g_{k} e^{i(\omega_{0}-\omega_{k})t} C_{k}(t)=E C_{N}(t).
\end{array}
\end{eqnarray}
Obtaining $C_{k}(t)$ from the first equation and substituting it into the rest ones gives the following set of $N$ integral equations
\begin{eqnarray}
\begin{array}{c}
  (E-\omega_{0}) C_{1}(t)=-\int_{0}^{\infty} \frac{J(\omega) d\omega}{E-\omega} \sum_{l=1}^{N} C_{l}(t),\\\\
  (E-\omega_{0}) C_{2}(t)=-\int_{0}^{\infty} \frac{J(\omega) d\omega}{E-\omega} \sum_{l=1}^{N} C_{l}(t),\\
  .\\
  .\\
  .\\
  (E-\omega_{0}) C_{N}(t)=-\int_{0}^{\infty} \frac{J(\omega) d\omega}{E-\omega} \sum_{l=1}^{N} C_{l}(t).
\end{array}
\end{eqnarray}
Consequently, by eliminating the amplitudes, these equations can be combined in a compact form as
\begin{eqnarray}
E=\omega_{0}-N \int_{0}^{\infty} \frac{J(\omega) d\omega}{E-\omega} \equiv y(E).
\end{eqnarray}
Solution of Eq. (14) highly depends on the particular choice of the spectral density of the reservoir.
It is clear that the existence of a bound state in the spectrum of Eq. (11) requires that Eq. (14) must have at least a real solution in the negative energy range, i.e. $E < 0$.
In general, as discussed in Ref. $\cite{Tong}$, existence of bound states in the spectrum of the total Hamiltonian depends on the fact that Eq. (14) must satisfy the condition $y(0) < 0$, otherwise formation of bound state is suppressed.
In Sec. V, the constructive role of the manipulation of the total system bound states on the speeding up of single-qubit evolution through additional qubits is demonstrated. In this regard, we examine the method by two concrete models such as Lorentzian and Ohmic spectral function for the reservoir structure.

\section{Quantum speed limit time}

Following the Ref. $\cite{Deffner}$, we give briefly the formal derivation of QSL time bound for the $1^{th}$ single-qubit open system in the presence of $N-1$ additional qubits corresponding to the model described in Sec. II. On the basis of geometric approach provided by the Bures angle $B(\rho_{10}, \rho_{1\tau})$ between the initial pure state $\rho_{10}=|\psi_{0}\rangle\langle\psi_{0}|$ where $|\psi_{0}\rangle$ is the state in (7) with $C_{l}(0)=0$ for $l=2, 3, ..., N$, and the target mixed state $\rho_{1\tau}$ defined as follows
\begin{eqnarray}
\mathcal{B}(\rho_{10},\rho_{1\tau})=\mathrm{arccos}\Big(\sqrt{\langle\psi_{0}|\rho_{1\tau}|\psi_{0}\rangle}\Big).
\end{eqnarray}
By taking time derivative of the Bures angle (15), the following inequality is easily obtained
\begin{eqnarray}
2\mathrm{cos}(\mathcal{B})\mathrm{sin}(\mathcal{B})\dot{\mathcal{B}}\leq|\langle\psi_{0}|\dot{\rho}_{1t}|\psi_{0}\rangle|.
\end{eqnarray}
The nonunitary dynamics for the considered $1^{th}$ qubit is described by the following exact master equation
\begin{eqnarray}
\dot{\rho}_{1t}=\Gamma(t)[2\hat{\sigma}^{-}_{1}\rho_{1t} \hat{\sigma}^{+}_{1}-\rho_{1t}\hat{\sigma}^{+}_{1} \hat{\sigma}^{-}_{1}-\hat{\sigma}^{+}_{1} \hat{\sigma}^{-}_{1}\rho_{1t}]:=\mathcal{L}_{t}(\rho_{1t}),
\end{eqnarray}
where $\Gamma(t)+i\Omega(t)=-\frac{\dot{C}_{1}(t)}{C_{1}(t)}$ and $C_{1}(t)$ is determined by the Eq. (9). By substituting Eq. (17) into Eq. (16), it is obtained
\begin{eqnarray}
2\mathrm{cos}(\mathcal{B})\mathrm{sin}(\mathcal{B})\dot{\mathcal{B}}\leq|\mathrm{tr}\{\mathcal{L}_{t}(\rho_{1t})\rho_{10}\}|.
\end{eqnarray}
Consequently, by employing the von Neumann trace inequality $\cite{Deffner}$, Eq. (18) takes the form
\begin{eqnarray}
2\mathrm{cos}(\mathcal{B})\mathrm{sin}(\mathcal{B})\dot{\mathcal{B}}\leq\|\mathcal{L}_{t}(\rho_{1t})\|_{\mathrm{op}},
\end{eqnarray}
where $\|.\|_{\mathrm{op}}$ denotes the operator norm. Integrating Eq. (19) over actual driving time $\tau$, we find
\begin{eqnarray}
\tau\geq\tau_{\mathrm{QSL}}=\frac{\mathrm{sin}^2\big[\mathcal{B}(\rho_{10},\rho_{1\tau})\big]}{\Lambda^{\mathrm{op}}_{\tau}},
\end{eqnarray}
where $\Lambda^{\mathrm{op}}_{\tau}=\frac{1}{\tau}\int_{0}^{\tau}dt\|\mathcal{L}_{t}(\rho_{1t})\|_{\mathrm{op}}$.
Suppose that the initial state of our described system (see section II) is $\rho_{1}(0)=|1\rangle \langle1|$, which means that the $1^{th}$ qubit is in the excited state
and $N-1$ additional qubits along with the reservoir are in their respective ground state.
So the QSL time bound for this qubit can be evaluated as

\begin{eqnarray}
\tau_{QSL}=\frac{\tau(1-|C_{1}(\tau)|^2)}{\int_{0}^{\tau} \partial_{t}|C_{1}(t)|^{2}dt},
\end{eqnarray}
where $C_{1}(t)$ can be calculated from Eq. (9). By Eq. (21), the QSL time bound for the $1^{th}$ qubit evolution not only depends on the spectral density of the reservoir but also on the existence of bound states in the spectrum of the total Hamiltonian which can be manipulated by including or excluding other qubits which will be discussed in the next section. We will take Lorentzian and Ohmic spectral density for the explicit solution of Eq. (9) where it can be solved analytically for the Lorentzian and numerically for the Ohmic one.

\section{Results}

\subsection{Lorentzian spectral density}

In this step, we attempt to demonstrate the predicted results explicitly by taking Lorentzian structure for the reservoir as
\begin{eqnarray}
  J(\omega)=\frac{1}{2\pi} \frac{\gamma_{0} \lambda^{2}}{(\omega-\omega_{0})^2+\lambda^{2}},
\end{eqnarray}
where the parameter $\lambda$ defines the spectral width of the coupling and $\gamma_{0}$ is the coupling strength of the system to the reservoir. It should be noted that finding an analytical criterion for the formation of bound state, i.e. $y(0) < 0$ (see Eq. (14)), is impossible for the case of Lorentzian reservoir so this condition is evaluated numerically. Fig. 1(a), shows the behaviors of energy spectrum of the total Hamiltonian in terms of the coupling strength $\gamma_{0}$. Clearly, formation of bound states in terms of $\gamma_{0}$ in the presence of additional qubits, happens faster relative to the absence of them. Also, for a given $\gamma_{0}$, as the number of qubits grows up the degree of boundedness becomes stronger.

In the next step, we are going to calculate the QSL time for the evolution of $1^{th}$ qubit and show how it can be controlled by manipulation of bound states through additional qubits. Fortunately, for the Lorentzian reservoir the probability amplitude $C_{1}(t)$ can be obtained analytically from Eq. (9). Explicitly, by considering that the $1^{th}$ qubit is only initially excited, its probability amplitude at time $t$ is calculated as
\begin{eqnarray}
  C_{1}(t)=G(t)C_{l}(0),
\end{eqnarray}
where
\begin{eqnarray}
  G(t)=\frac{N-1}{N}+\frac{e^{-\lambda t/2}}{N} \bigg(\mathrm{cosh}{(\frac{Dt}{2})}+\frac{\lambda}{D} \mathrm{sinh}{(\frac{Dt}{2})}\bigg),
\end{eqnarray}
with $D=\sqrt{\lambda^{2}-2\gamma_{0} \lambda N}$. Consequently, the QSL time bound for the $1^{th}$ qubit evolution is evaluated from Eq. (21).

In Fig. 1(b), the behavior of the QSL time bound, by considering the other non-interacting qubits, has been sketched in terms of $\gamma_{0}$. Interestingly, increasing the number of additional qubits leads more decrement of the QSL time bound. This shows that we can speedup arbitrarily the evolution of the $1^{th}$ qubit by making stronger bound states in the spectrum of the total system through the including of other qubits into the reservoir.

\subsection{Ohmic spectral density}

In order to confirm the performance of the speedup mechanism, we proceed further by taking the Ohmic spectral density as
\begin{eqnarray}
  J(\omega)=\frac{\gamma}{2\pi} \omega e^{-\frac{\omega}{\omega_{c}}}.
\end{eqnarray}
where $\omega_{c}$ is the cut-off frequency and $\gamma$ is a dimensionless coupling constant.
As discussed in section III, a bound state is formed if the
condition $y(0) < 0$ is satisfied, therefore, for the Ohmic spectral density we have
\begin{eqnarray}
  y(0) < 0 \rightarrow \gamma > \frac{2\pi\omega_{0}}{N \omega_{c}}=\gamma_{c}.
\end{eqnarray}
From Eq. (26), this condition can be well-satisfied by inserting other qubits into the reservoir.
It is interesting to note that , as $N \rightarrow \infty$, the critical value $\gamma_{c}$ reaches to zero,
which means that there are no restrictions on the formation of bound states.
Fig. 2(a) shows the negative energy spectrum of the total Hamiltonian in terms of $\gamma$ for some set of additional qubits, when the spectral density of the reservoir is Ohmic. As for the Lorentzian case, stronger bound states are created by inserting more additional qubits into the reservoir. Unfortunately, in contrast to the Lorentzian spectral density, there is no analytical solution for the integro-differential equation (9). Therefore, QSL time bound is obtained numerically, as depicted in Fig. 2(b). It is also confirmed that by attendance of other qubits into the reservoir, the QSL time bound of the single-qubit evolution is decreased again. It is concluded that, for both reservoirs, the quantum evolution speedup is well-controlled by manipulating of bound state spectrum of the whole system through additional qubits.

\section{Conclusions}
In summary, the mechanism of well-controlled quantum speedup of single-qubit evolution has been explored
on the basis of creating bound states for the total system through the inserting non-interacting other qubits.
Although, as was observed, the speed of evolution in the case of Ohmic reservoir is generally higher than the Lorentzian one, However, it seems that in the presence of much more additional qubits the speed of evolution of the system takes the same value, irrespective to the structure of the reservoir.

\newpage
Fig. 1. (a) The negative energy spectrum of the system-reservoir in terms of the coupling strength $\gamma_{0}$ (in units of $\omega_{0}$) for the reservoir with Lorentizan spectral density.
(b) The QSL time bound of the single-qubit evolution in terms of $\gamma_{0}$ (in units of $\omega_{0}$) for $\lambda=1$ (in units of $\omega_{0}$) and $\tau=10$.
\begin{figure}
        \qquad \qquad\qquad\qquad \qquad a \qquad\qquad \quad\qquad\qquad\qquad\qquad\qquad\qquad b\\{
        \includegraphics[width=3in]{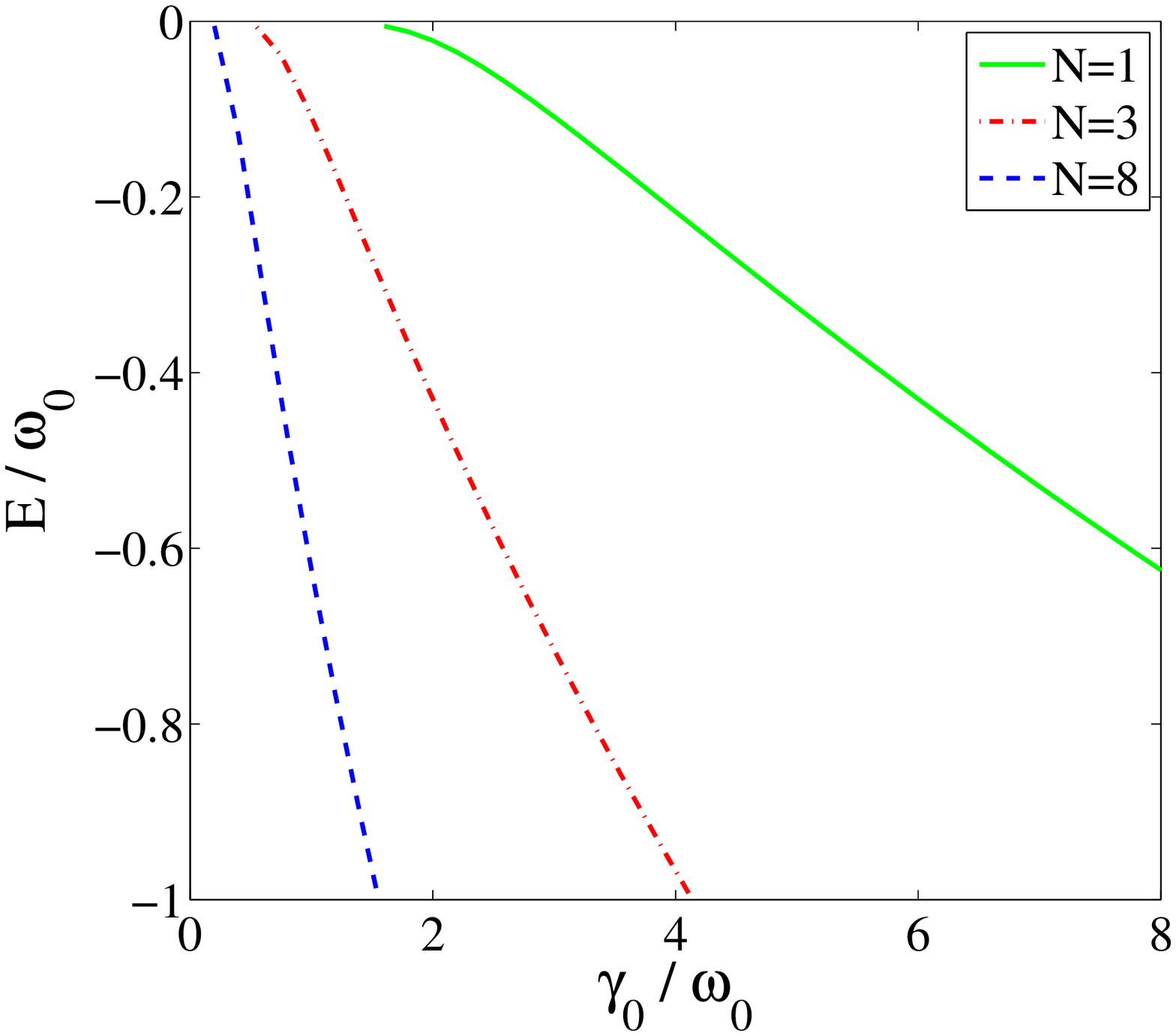}
        \label{fig:first_sub}
    }{
        \includegraphics[width=3in]{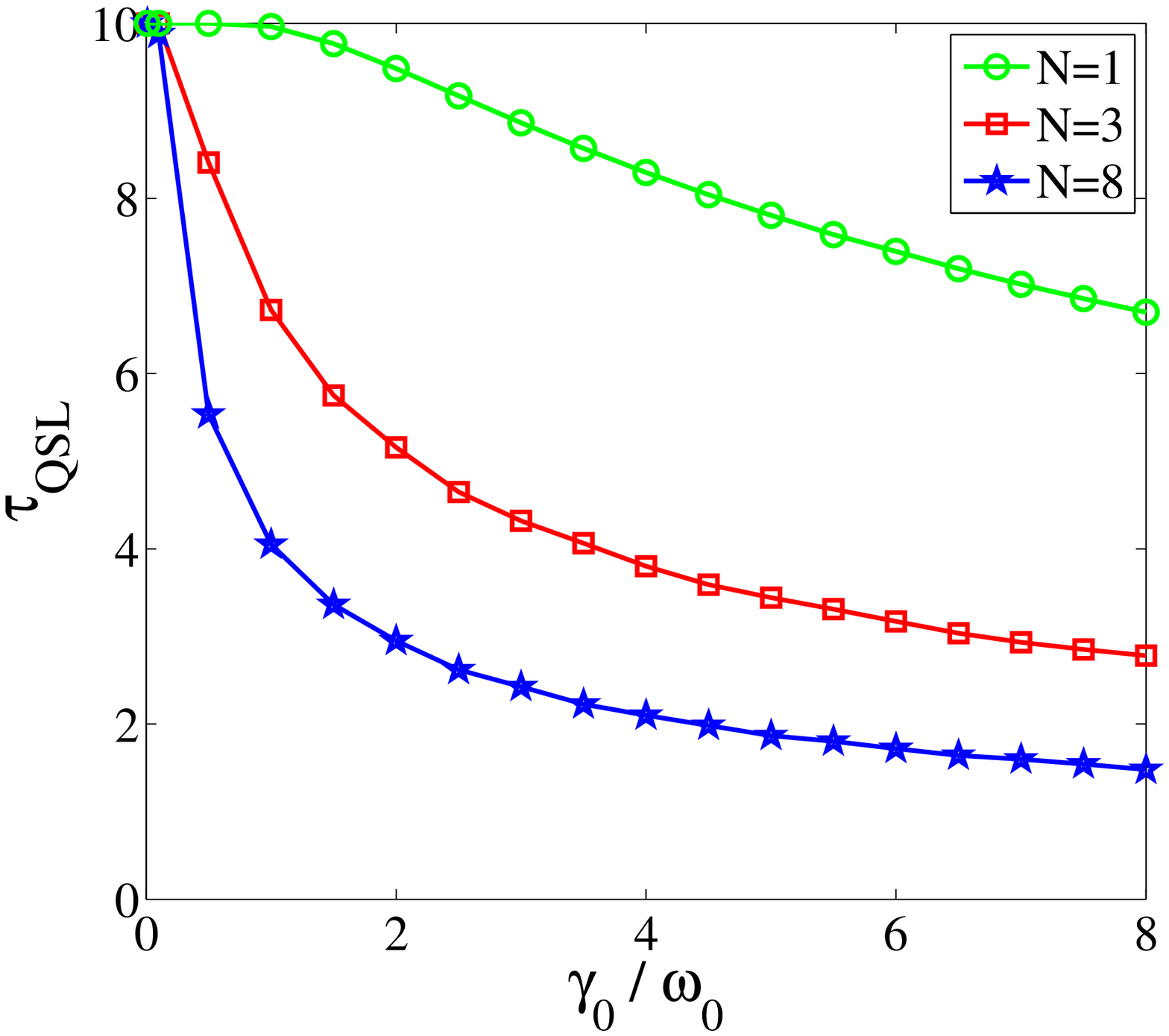}
        \label{fig:second_sub}
    }
    \caption{}
    \end{figure}

\newpage
Fig. 2. (a) The negative energy spectrum of the system-reservoir in terms of the coupling constant $\gamma$ (in units of $\omega_{0}$) for the reservoir with Ohmic spectral density.
(b) The QSL time bound of the single-qubit evolution in terms of $\gamma$ (in units of $\omega_{0}$) for $\lambda=1$ (in units of $\omega_{0}$) and $\tau=10$.
\begin{figure}
        \qquad \qquad\qquad\qquad \qquad a \qquad\qquad \quad\qquad\qquad\qquad\qquad\qquad\qquad b\\{
        \includegraphics[width=3in]{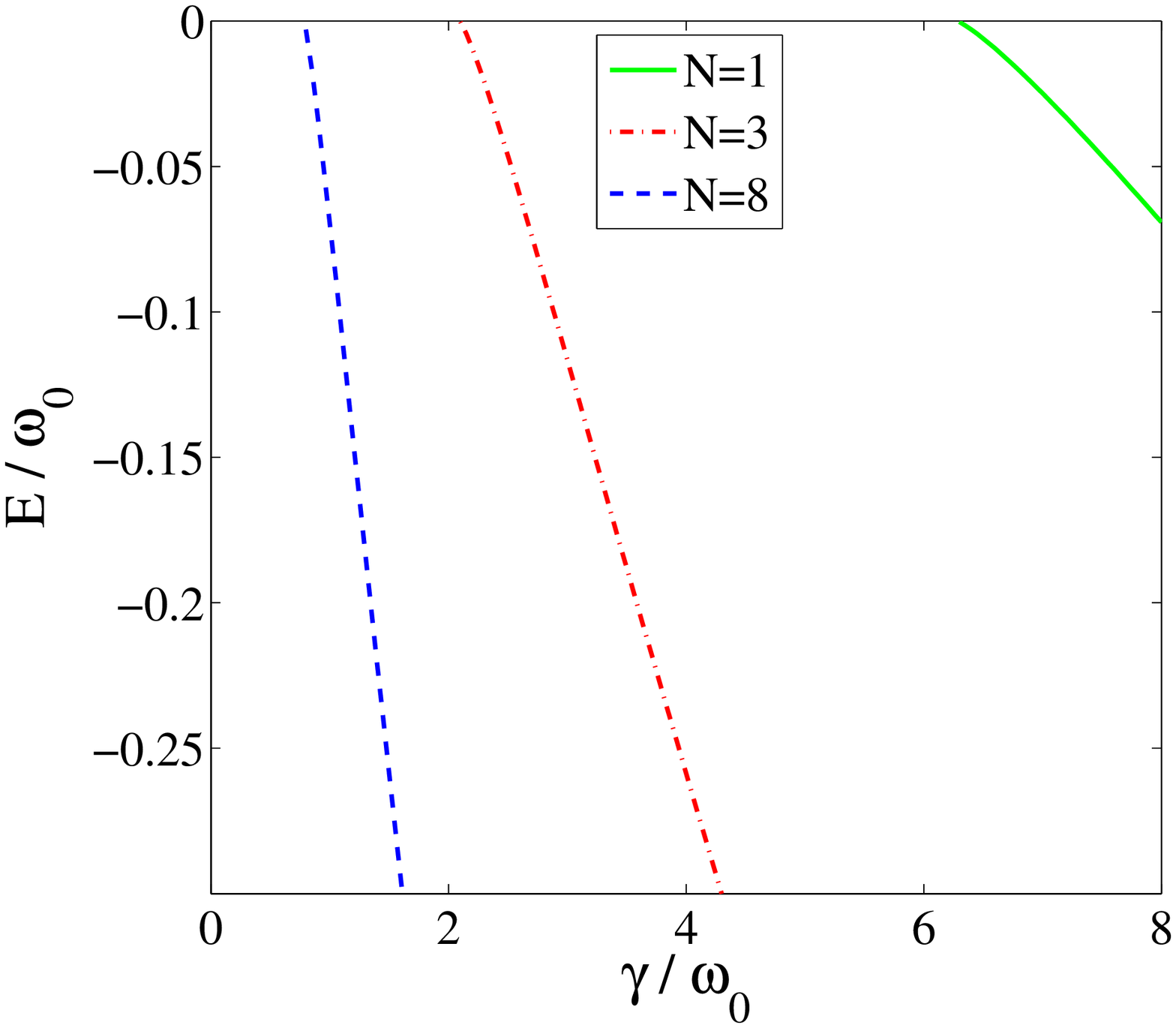}
        \label{fig:first_sub}
    }{
        \includegraphics[width=3in]{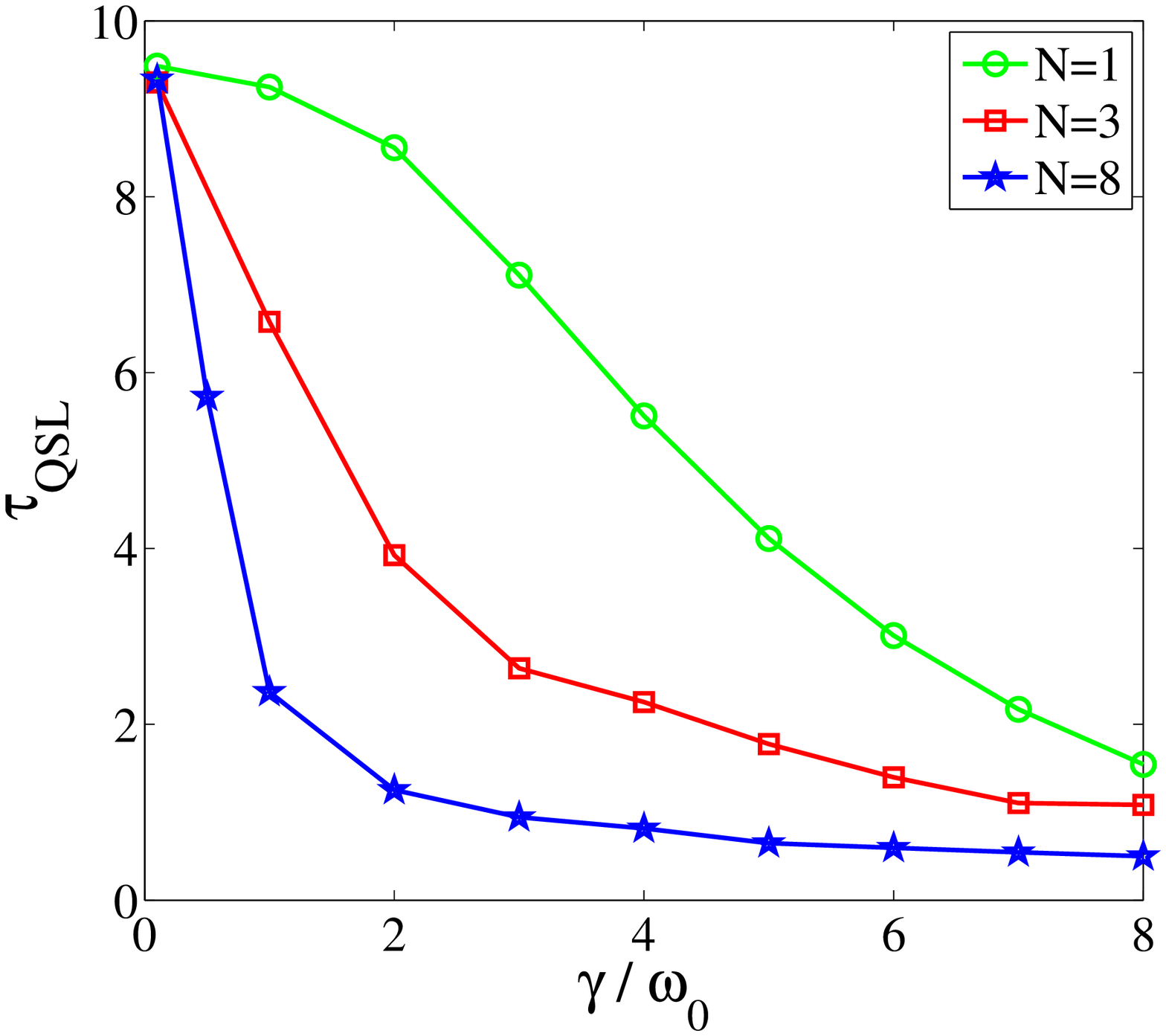}
        \label{fig:second_sub}
    }
    \caption{}
    \end{figure}

\end{document}